\documentclass[aip,jap,reprint,showpacs]{revtex4-1}
\usepackage{graphicx}
\usepackage{dcolumn}
\usepackage{bm}
\usepackage{amsmath}%
\usepackage{sidecap}
\usepackage[top=1.7cm, bottom=1.6cm, left=1.5cm, right=1.5cm]{geometry}
\usepackage{xcolor}

\begin{document}

\title{\color{blue} Disorder-free localization around the conduction band edge of crossing and kinked silicon nanowires}

\author{\"{U}mit Kele\c{s}}
\author{Asl{\i} \c{C}akan}
\author{Ceyhun Bulutay}
\email{bulutay@fen.bilkent.edu.tr}
\affiliation{Department of Physics, Bilkent University, Bilkent, Ankara 06800, Turkey}

\date{\today}

\begin{abstract}
\fontsize{10}{10}\vskip3pt\noindent \hskip0cm\begin{minipage}{14cm} \parindent.2in \noindent 
We explore ballistic regime quantum transport characteristics of oxide-embedded crossing
and kinked silicon nanowires (NWs) within a large-scale empirical pseudopotential electronic
structure framework, coupled to the Kubo-Greenwood transport analysis. A real-space wave function study
is undertaken and the outcomes are interpreted together with the findings of ballistic transport calculations.
This reveals that ballistic transport edge lies tens to hundreds of millielectron volts above the lowest
unoccupied molecular orbital, with a substantial number of localized states appearing in between, as well as above the former. We show that these localized states are not due to the oxide interface, but rather core silicon-derived.
They manifest the wave nature of electrons brought to foreground by the reflections originating from NW junctions and bends. Hence, we show that the crossings and kinks of even ultraclean Si NWs possess a conduction
band tail without a recourse to atomistic disorder. \hfil \end{minipage}\normalsize\vskip12pt

\end{abstract}




\maketitle

\section{Introduction}
As the fabrication techniques for silicon nanowire (NW) polymorphs are relentlessly
advancing,\cite{morales1998,holmes2000,cui2001wire,ma2003,wu2004,martinez10,ozdemir11,barbagiovanni2014}
the latest accomplishment has been their assembly into higher dimensional
complex architectures. With this aspiration, a new class of nanostructures emerged,
including L-shaped\cite{westwater1997,tian2009single} and kinked\cite{westwater1997}
as well as hyper-branching\cite{wang2004,jiang2011} and networking Si NWs.\cite{mulazimoglu2013}
Among these, kinked NWs (KNWs) in particular have received special interest owing to their rich
geometries including U-, V-, and W-shaped varieties.\cite{tian2009single,tian2010kn,cook11,xu2013}
They have potential application areas ranging from biological and chemical sensing,
nanoscale photon detection to three-dimensional recording from within living cells
and tissue.\cite{jiang12,qing14,xu14}

These NW-based superstructures in their ultraclean forms can be regarded as the electron
waveguide components making up a quantum network, once theoretically
envisioned.\cite{sheng97} As a matter of fact, the field has a rich
background.\cite{ferry} The dawn of the ballistic transport era in low-dimensional semiconductors
can be marked by the quantum interference transistor proposal of Datta and coworkers that was
based on exerting a phase difference between the two electron channels.\cite{datta86}
This was followed by experimental reports on GaAs/AlGaAs heterojunctions demonstrating resistive
increase when the current path encounters a bend,\cite{timp88} and also resistance oscillations
on T-shaped electron channel when the stub length is electrically tuned.\cite{aihara93}
More recently, as observed in a 500~nm-long silicon NW, quantum interference effects cause an
oscillatory pattern in conductance as well as a shift in conduction threshold depending on the
quantization energy within the wires.\cite{tilke03} Despite the intense efforts, due to
imposing challenges in avoiding diffusive transport channels, the elicitation of experimental
fingerprints endorsing quantum interference effects remain still very subtle.

Without any doubt, any theoretical insight would be invaluable to better understand the rich physics
and application prospects of these structures such as silicon-based crossing NWs (CNWs) and KNWs.
If one sets the objective to characterize the full mosaic of localized to delocalized states,
then it becomes imperative to acquire atomistic resolution, and therefore the methodology needs to
surpass the effective mass, $\textbf{k}\cdot\textbf{p}$, or transfer matrix-like continuum
models.\cite{weisshaar91,shao94,sengupta05,cook11}
On the other hand, due to large system sizes in CNWs and KNWs,
on the order of ten thousands of atoms including the silica embedding host matrix, they are not
amenable by the current state-of-the-art \textit{ab initio} techniques.\cite{martin2004} Because
of these constraints, only a small number of realistic-size atomistic NW studies
exist,\cite{neophytou2008,persson08,kim11} leaving the CNWs and KNWs as an unchartered territory.

This work aims to provide a computational exploration of the full scale, from localized
to semi-extended up to conducting characteristics of electronic states in Si CNWs
and KNWs. We employ an empirical pseudopotential-based atomistic electronic structure solved
using the linear combination of bulk bands (LCBB) method.\cite{lcbb1997,lcbb1999}
Very recently, we have successfully applied this technique on the bandgaps and band edge alignments of
Si NWs and NW networks.\cite{keles} In conjunction with the quantum ballistic
transport calculations utilizing the Kubo-Greenwood (KG) formula, we undertake an isosurface analysis
of wave function distributions for the CNWs and KNWs.
Our main finding is the identification of a conduction band tail of up to several hundred
millielectron volts in span. Unlike the well-known Urbach tail associated with an atomistic
disorder,\cite{urbach53,pan08} it stems from the electron wave localization within the
junction and kink sections of ultraclean Si CNWs and KNWs.

The temperature range relevant to this ballistic regime is a few Kelvins, even though we shall be
presenting zero-temperature results. Another noteworthy point is the strain which can in general
play a significant role on the electronic properties of Si NWs.\cite{leu2008,shiri08,niquet2012}
Yet, a uniform strain (radial or axial, compressive or tensile) retains the extensions of
states throughout the NW and does not lead to localized states.\cite{leu2008,wu2009}
In contrast, inhomogeneous strain can promote the localization behavior in NWs; in the
same vein, interface roughness can induce disorder-originated localized states.\cite{svizhenko2007,persson08,Lherbier08}
In this work, targeting atomistic disorder-free localization, the strain, as well as the interface
relaxation effects, are left out of the scope, to unambiguously account for the effects of
wire topology. Nevertheless, including the strain and interface disorder is expected to
further intensify and enforce the present localization behavior.

The paper is organized as follows: in Sec.~II we present the general theoretical framework of our
electronic structure and transport analysis. In Sec.~III we provide our results on CNWs and KNWs,
and conclude in Sec.~IV. In the interest of a lucid presentation, we defer some technicalities
to two Appendixes. The first one provides the essential details on the LCBB technique,
followed by the details on the atomic pseudopotentials employed in this work.

\section{Theory}
For the electronic structure calculations of large-scale
nanostructure systems the LCBB method has been
introduced,\cite{lcbb1997,lcbb1999} with the distinction to solve
the atomistic pseudopotential-based Hamiltonian using a basis set
formed by the bulk bands of the constituent materials. The virtue
of the method lies in the nanostructure sizes accessible with a
reasonable computational budget.  Within a 3-dimensional (3D)
supercell approach the complex NW structures are embedded into an
oxide matrix to passivate the dangling bonds and to prevent the
interactions between the structures and their images in
neighboring cells. The nanostructures considered in this work
contain about 1500 core Si atoms, and including the host medium
the supercells contain more than 10\,000 atoms. Perfect
crystalline order is considered as no structural relaxation is
performed. Being interested in conduction states, we neglect the
spin-orbit interaction which is mainly influential on the valence
states. Details of the LCBB formalism, description of oxide passivation
and the employed pseudopotential form factors
can be found in Appendixes A and B.

To study the linear-response regime ballistic quantum transport characteristics of the nanostructures,
the KG-formula\cite{kubo1957,greenwood1958} is applied as a postprocess over their
electronic structures. The zero-temperature dc-limit KG-formula for conductance is given as
\begin{equation} \label{kuboformula}
\begin{split}
G_{\alpha \beta} (E_F) = & g_s \frac{\pi \hbar e^2}{m_o^2 L_{sc}^2}
\sum_{n,n^{\prime}} \langle n^{\prime} \vert P_{\alpha} \vert n
\rangle \langle n \vert P_{\beta} \vert n^{\prime} \rangle \\ & \times \delta(E_F-E_n) \delta(E_F-E_{n^{\prime}}) \, ,
\end{split}
\end{equation}
where $\alpha , \beta$ are the Cartesian indices, $g_s$ is the spin degeneracy, $m_o$ is the free electron mass,
and $L_{sc}$ is the length of the computational supercell along the chosen transport direction,
$P$ is the momentum operator and $E_n$ is the energy eigenvalue of the eigenstate $\vert n \rangle$.\cite{akkermans}
The KG-formula establishes the relation between the electronic structure of a system and its conduction properties,
as a function of Fermi energy $E_F$. In our implementation, the Dirac delta functions of Eq.~(\ref{kuboformula})
are broadened by the Gaussian function with a typical standard deviation of 5\,meV.

\begin{figure}[t]
\centering
\includegraphics[width=8.5cm]{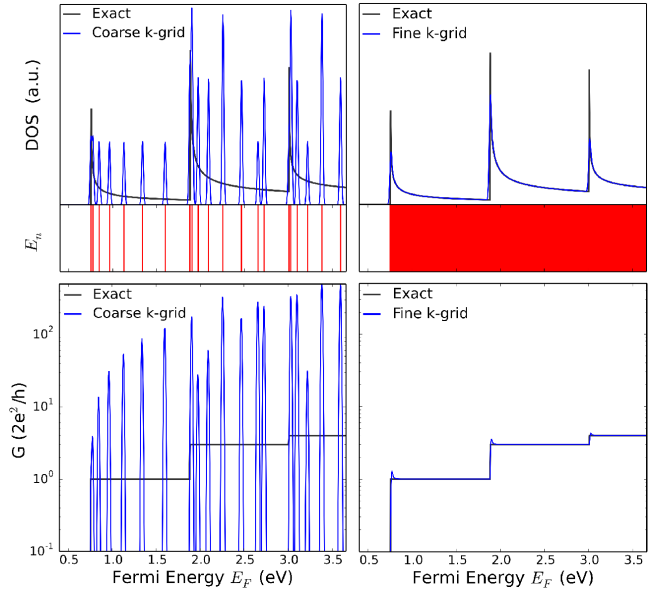}
\caption{ \footnotesize Demonstration of the coalescence of isolated conductance peaks with $k$-grid resolution
to ideal steps for a representative square cross-section hard-wall NW.
Logarithmic scale is used for conductance to accommodate all the peaks.
$E_n$ panels show the energy spectra for
the associated $k$-grid.} \label{comp_anlytic_fig}
\end{figure}

It is worthwhile to discuss some technical aspects of the implementation of KG-formula.\cite{imry,thesis}
Fig.~\ref{comp_anlytic_fig} shows two representative cases for a free-electron model 1~nm$\times$1~nm
cross-section hard-wall NW. On the left column, the coarse-grid case is displayed in which the sparse energy
spectrum ends up with the Dirac delta function-derived isolated peaks both for the density of states (DOS) and conductance.
The performance of the griding depends on $\Delta E/ \gamma$, i.e., the ratio of the energy difference of
successive $k$-points, $\Delta E$, to the standard deviation of Gaussian broadening, $\gamma$.
In the coarse case, the ratio is about 5 at the band edge and further increases through the band dispersion.
Since the energy difference between successive conductance steps is larger than the energy broadening,
the isolated peaks of both DOS and conductance do not coalesce to
reproduce the well-known analytically expected 1D DOS and
quantized conduction behavior, a signature of the 1D ballistic transport.\cite{vanwees88,ferry}
Our tests indicate that as the ratio drops below 0.02, the coalescence of isolated peaks starts to yield reasonable
DOS and conductance shapes. On the right column of representative fine-grid case, by increasing the number of wave vector, \textbf{k}-points in the energy dispersion curve, the spectrum is made dense enough so that the ratio is set about 0.005 to obtain the analytical behaviors. Hence, this figure signifies the necessity for a dense energy spectrum in achieving ideal quantized conductance steps through the KG-formula.

\begin{figure}[t]
\centering
\includegraphics[width=8.5cm]{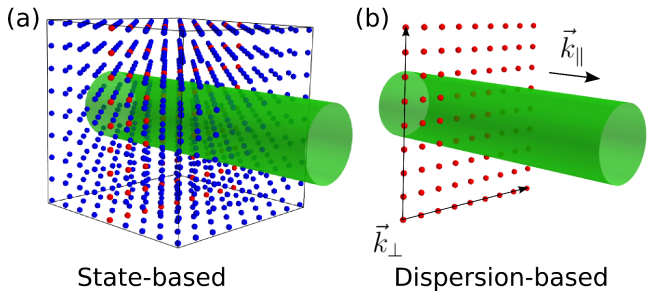}
\caption{ \footnotesize Schematic illustration of (a)~3D and (b)~2D
k-grids. The tube indicates the alignment of considered dispersion direction.  The planar
$k$-grid $\vec{k}_{\perp}$ is shifted gradually along the dispersion direction
$\vec{k}_{\parallel}$.} \label{kgrid_fig}
\end{figure}

Notwithstanding, a typical LCBB calculation based on a 3D \textbf{k}-grid (see Fig.~\ref{kgrid_fig}\,(a) and also Appendix~A)
yields relatively sparse discrete energies.
Obviously, the energy spectrum can be made denser  (i.e., $\Delta E/ \gamma$ reduced) by
increasing the number of \textbf{k}-points in the 3D grid (i.e., in the expansion basis),
however it becomes computationally quite demanding.
Instead, regarding the \textbf{k}-space, we construct a dense $\textbf{k}_{\perp}$-grid \textit{over the plane} perpendicular to the KNW
axis or to the axis of one of the crossing wires of a
CNW, and obtain the energy dispersion relation $E_n(k_{\parallel})$ of subbands by shifting this plane along that
axis of $\textbf{k}_{\parallel}$ (see Fig.~\ref{kgrid_fig}\,(b) and also Appendix~A). Once a reasonable dispersion is obtained,
by further interpolating the energy spectrum the
KG-formula can be applied to disclose the conductance steps.
We call this the \textit{dispersion-based} approach. It should be noted that in the 3D $k$-grid case which
we shall name as the \textit{state-based} approach, the interpolation is not possible since the 3D eigenvalue problem
yields only state energies, $E_n$ without revealing their $k$-dependence. Thus, KG-formula inevitably
produces isolated conductance peaks for this state-based case.

For clean single NWs having no localized states, the dispersion- and state-based approaches yield the same
energy spectrum---once constructed with the same number of \textbf{k}-points.  However, in the case of
CNWs and KNWs the findings of these two approaches deviate: while the state-based approach is able to
capture both the states having localized or extended characters, the dispersion-based approach finds
only the extended-dispersive states. As we mentioned above, the state-based approach at full performance
becomes computationally expensive, so we apply a \textit{two-pass} approach, where we get the extended states
through the dispersion-based approach, followed by a reasonable-budget state-based approach to single out the
localized states and to mark the ballistic transport edge (BTE).

\vspace{-0.6cm}

\section{Results}
\vspace{-0.4cm}
\subsection{Crossing Nanowires}
\vspace{-0.3cm}

\begin{figure}[h]
\centering
\includegraphics[width=8.5cm]{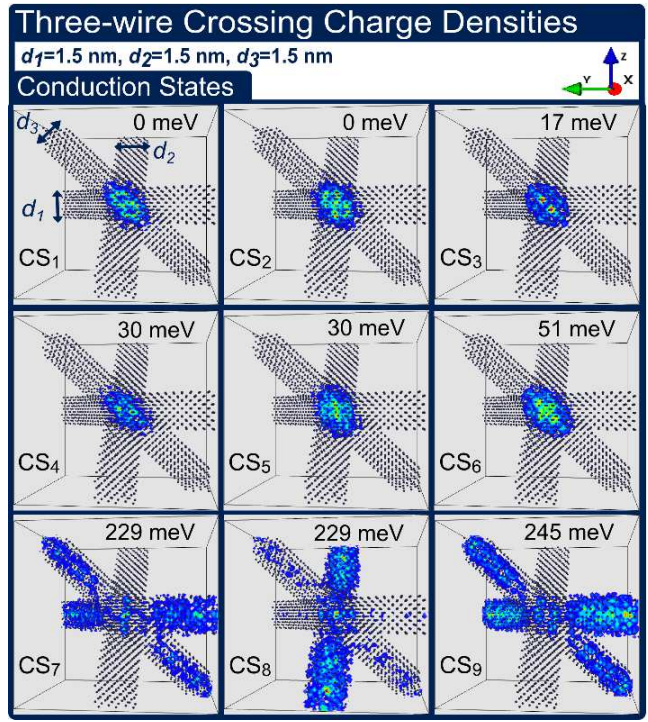}
\caption{ \footnotesize Isosurfaces for charge densities corresponding to the
first nine states above the conduction band edge for the three-wire
crossing with wire orientations along
$\langle110\rangle$ directions and diameters of all 1.5\,nm.
For clarity, the surrounding oxide atoms are not shown. In this and following figures, LUMO energy is set to zero,
and the isosurfaces enclose densities from 100\% to 10\% of their respective maxima.}
\label{iso111}
\end{figure}

We start our transport analysis targeting the low-lying conduction band edge states with
the CNWs made up of three crossing 1.5\,nm diameter Si NWs oriented along the
$\langle 110 \rangle$ directions. We choose the $\langle 110 \rangle$ Si NWs since it is the major
growth orientation within the sub-10-nm diameter size regime.\cite{wu2004} In Fig.\,\ref{iso111},
we present the isosurfaces of carrier densities (wave function modulus squares) of the nine
lowest electron states, where the first one, CS$_{\rm 1}$ corresponds to the lowest unoccupied
molecular orbital (LUMO).\cite{note1} 
Among them the first six states, CS$_{\rm 1}$-CS$_{\rm 6}$
are observed to be highly localized at the junction of the CNW and diminishes sharply beyond the junction.
Energetically, they nest in a window of 51\,meV with the individual
splittings being governed by the CNW as well as the atomistic point group symmetries as in Si
nanocrystals.\cite{bulutay2007} Separated from this group by about 200\,meV, another set of
states exists, this time having extended wave functions spread to some portions of the network.
As the figure demonstrates, all the states are Si core-derived, with no interface trapping or penetration to the oxide
matrix.\cite{note1}

\begin{figure}[ht]
\centering
\includegraphics[width=8.5cm]{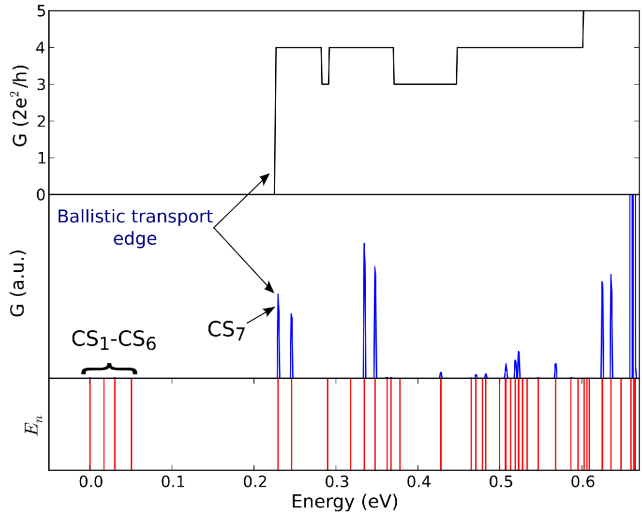}
\caption{ \footnotesize The calculated ballistic electron conductances and
energy spectrum of 1.5\,nm equal-diameter CNW.
Top: The conductance through the 1D channel of
[110] NW via the dispersion-based approach.
Center: The transport along the [110] direction within state-based approach.
Bottom: The energy spectra belonging to the center panel.
\label{1n5x3_T_fig} }
\end{figure}

To confirm our isosurface-based interpretations, we compare them with those
of the quantum ballistic transport results using the KG-formula.
Specifically, we apply the aforementioned two-pass treatment to harness their complementary
features to the CNW of Fig.\,\ref{iso111}. On the top panel of Fig.\,\ref{1n5x3_T_fig},
the dispersion-based approach is displayed where the quantized conductance
steps are observed. In the center panel,
state-based conduction results are presented. Since the Dirac delta-originated
peak values depend on the broadening parameters, only their relative amplitudes and sparsity
need to be considered. The energy spectrum underlying the state-based approach is given
on the bottom panel. All these results corroborate the isosurface analysis
of Fig.\,\ref{iso111}. Both the dispersion- and state-based approaches pinpoint to the
same energy level as the BTE, which is the energy level of CS$_{\rm 7}$ of Fig.\,\ref{iso111}.
It needs to be mentioned that in CNWs we have spotted localized states also \textit{above} the BTE.\cite{thesis}

Another important remark is that the BTE for CNW is very close to the conduction band minimum (and BTE) of an \textit{isolated}
NW of diameter of 1.5\,nm,\cite{keles} which affirms that indeed the geometric discontinuity of
the junction region in CNW is responsible for the additional localized states below BTE.
These findings point out that the quantum interference effects play a prominent role at the
crossing regions. Any discontinuity over the core material silicon not only excites higher order evanscent
modes\cite{weisshaar91,sols1989} but also results in reflections depending on the state energy and leads to standing waves.

\begin{figure}[b]
\centering
\includegraphics[width=8.5cm]{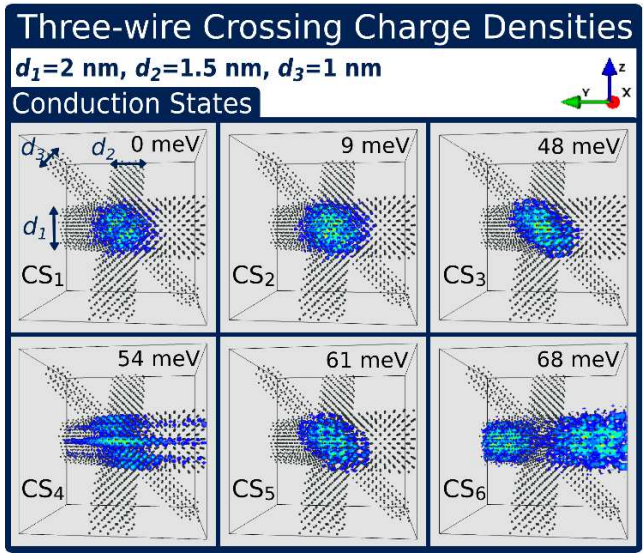}
\caption{\label{sup_iso111}  \footnotesize Isosurfaces for charge densities
corresponding to the low-lying conduction band
states for a $\langle110\rangle$ three-wire crossing with wire
diameters of 1~nm, 1.5~nm, and 2~nm.}
\end{figure}

\begin{figure}[ht]
\centering
\includegraphics[width=8.5cm]{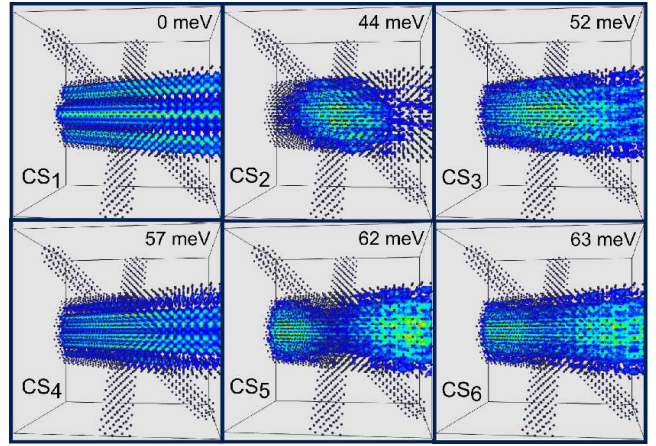}
\caption{\label{sup_iso3}  \footnotesize Isosurfaces for charge densities
corresponding to the low-lying conduction band
states for a $\langle110\rangle$ three-wire crossing with wire
diameters of 1~nm, 1~nm, and 3~nm.
Carrier densities extend along the widest crossing wire.}
\end{figure}

This situation gains a further dimension in the case of CNWs with \textit{all-different diameters}.
In Fig.\,\ref{sup_iso111}, we present the isosurfaces for charge densities of a three-wire
crossing with diameters of 1\,nm, 1.5\,nm, and 2\,nm. We observe that as the difference among the diameters
of crossing wires become larger, the electron states at the band edge delocalize, in particular,
along the thickest NW. Here, similar to the equal-diameter CNW of Fig.\,\ref{iso111}, the energy of the first
extended state CS$_{\rm 4}$ nearly coincides with the LUMO energy of an isolated 2~nm Si NW.\cite{keles}
In Fig.\,\ref{sup_iso3}, the isosurfaces are given
for the case of wire diameters of 1~nm, 1~nm, and 3~nm.
Concerning the LUMO, beyond a diameter contrast, the CNW behaves as isolated NW arrays, as if
the other thinner wires are not present. For the higher-lying states, while still
extending along the thickest wire, they get more sensitive to the thinner branches.
These figures show that the formation and profile of a localized/extended state depends on the relative
diameters of the NWs.

\subsection{Kinked Nanowires}
\begin{figure}[ht]
\centering
\includegraphics[width=8.5cm]{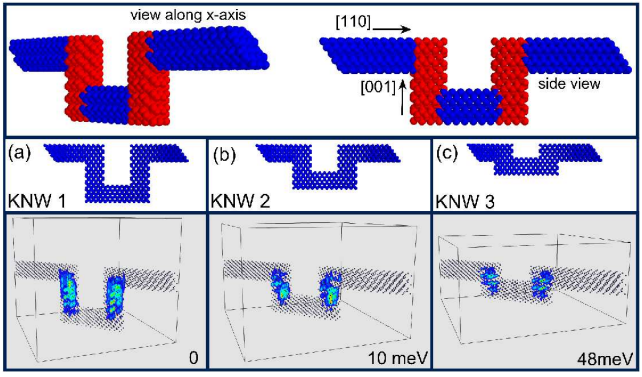}
\caption{ \footnotesize Top: A representative KNW depicted from two aspects.
The blue segments are along the [110] direction and the red ones are along [001]
direction. Bottom: Isosurfaces for charge densities of LUMOs of three
different shape KNWs. Their relative energies are also included.} \label{kinked_comp}
\end{figure}

Next, we examine KNWs, here one can benefit from various opportunities for the
electronic structure engineering by tuning the diameter and lengths of NW segments or
by changing the wire crystallographic orientation. Fig.\,\ref{kinked_comp} illustrates
the schematic view of a U-shaped KNW where the blue (red) segments are aligned in the
[110] ([001]) direction. On the same figure, we specify the variation of LUMO
energies as the length of the [001]-aligned segment is decreased. As observed from this
figure LUMO is particularly localized to these sections of the KNW forming a standing wave
along the [001] wires. Therefore, in agreement with the quantum size effect, the LUMO energy
increases as the length of [001] segments decrease.

\begin{figure}[b]
\centering
\includegraphics[width=8.5cm]{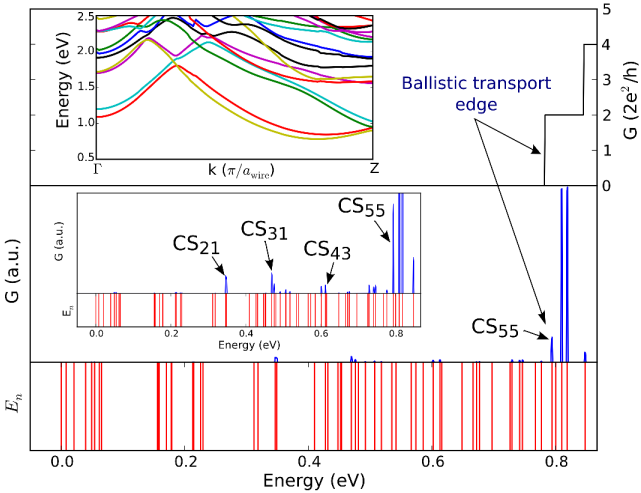}
\caption{\footnotesize  The calculated ballistic electron conductances and
energy spectrum of a 1.5\,nm diameter KNW.
Top: Conductance steps using the dispersion-based approach with the corresponding dispersion
being given in the inset. Center: The transport along the [110] direction within the
state-based approach, also zoomed in the inset.
Bottom: The energy spectra belonging to the center panel. \label{thelast}}
\end{figure}

Focusing on the KNW-3 sample of Fig.\,\ref{kinked_comp}, on the top panel of
Fig.\,\ref{thelast}, we first subject the KNW to the dispersion-based approach
which reproduces the conductance steps, i.e., NW-like dispersive states, starting from
the BTE about 0.8\,eV above LUMO. On the center panel, we present the conductance
values calculated by state-based approach.
Note that the onset of the major peaks is very close to the BTE indicated on the top panel.

\begin{figure}[t]
\centering
\includegraphics[width=8.5cm]{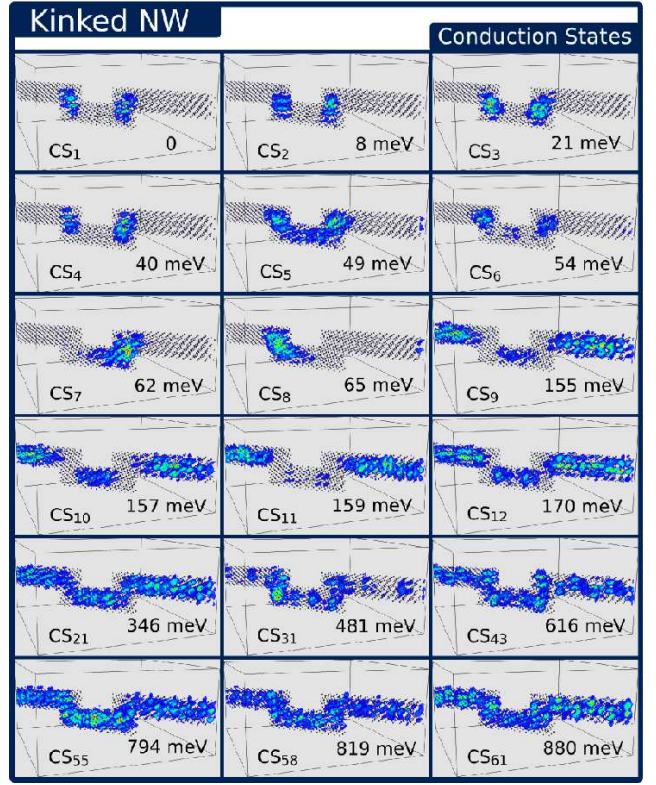}
\caption{ \footnotesize Isosurfaces for charge densities
corresponding to several conduction states (CS) of a KNW.} \label{kinked_states}
\end{figure}

We can classify the states in Fig.\,\ref{thelast} according to the relative values of their
conductance peaks: (S1) totally localized-bound states, not contributing to the conductance,
(S2) quasi-bound states, having low conductance peaks and energetically closer to BTE than to
LUMO, and (S3) NW-like extended states which have higher conductance peaks and lie beyond the BTE.
As a matter of fact these S3 states are responsible for the full 1D dispersion and contribute
to the conductance steps. Dispersion-based approach is not able to
capture S1 and S2 states, but identifies the S3 states.
The validity of this classification becomes more founded by correlating
these observations with the isosurface charge density profiles as presented in Fig.\,\ref{kinked_states}
for a selection of conduction states. Here, the first 12 states (CS$_{\rm 1}$-CS$_{\rm 12}$)
belong to S1-type states, indicating localized characteristics. The representative states
CS$_{\rm 21}$, CS$_{\rm 31}$, and CS$_{\rm 43}$ are S2-type also marked on Fig.\,\ref{thelast}.
As observed in Fig.\,\ref{kinked_states}, their isosurfaces also display an extended behavior.
Rest of the states given in Fig.\,\ref{kinked_states} are lying at or beyond the BTE,
exhibiting NW-like extensions (S3-type).

\vspace{-0.5cm}

\section{Conclusions}
In conclusion, combining wave function analysis with the ballistic transport
results we map out the localization behavior of silicon CNWs and KNWs.
We find that the BTE can be up to hundreds of millielectron volts above the LUMO, filled in
between with a substantial number of localized states. Their 3D wave functions unambigiously display
localization around the NW geometric discontinuity regions, such as junctions and bends which excite
higher-order evanscent modes and incubate probability amplitude interferences.
In contrast to the states close to LUMO, those in the
vicinity of BTE possess a proto-conducting character which might contribute to transport through an
assisted process, like phonons. Thus, via a realistic atomistic electronic structure analysis,
we show that ultraclean Si NW architectures such as CNWs and KNWs should manifest a
conduction band tail that exemplifies a disorder-free localization.
\vspace{-0.5cm}

\begin{acknowledgments}
This work has been supported by The Scientific and Technological Research Council of Turkey
(T\"{U}B\.{I}TAK) with Project No.~109R037.
\end{acknowledgments}

\appendix

\section*{Appendix A: LCBB Formalism}
\vspace{-0.3cm}
In the LCBB method,\cite{lcbb1997,lcbb1999,thesis} the $j^{th}$ state of the nanostructure
can be expressed as the
linear combination of bulk Bloch bands
\begin{equation*} \label{lcbb_01}
\Psi_{j}(\vec{r})=  \frac{1}{\sqrt{N}} \sum_{n,\vec{k},\sigma}
C^{\sigma}_{n,\vec{k},j}\,
u^{\sigma}_{n,\vec{k}}(\vec{r}) e^{i\vec{k}\cdot~\vec{r}} ,
\end{equation*}
where $N$ is the number of bulk primitive unit cells within the
large supercell of the nanostructure, $n$ is the bulk band index,
$\vec{k}$ is the wave vector sampling the BZ of the
underlying lattice, and $\sigma$ labels the constituent materials,
i.e., core or embedding media.
The periodic part $u^{\sigma}_{n,\vec{k}}(\vec{r})$ of the bulk
Bloch function can be expanded by the plane waves as
\begin{equation*}
 u^{\sigma}_{n,\vec{k}}(\vec{r})=\frac{1}{\sqrt{\Omega_o}}\sum_{\vec{G}}^{N_G}
 B^{\sigma}_{n\vec{k}}(\vec{G})e^{i\vec{G}\cdot\vec{r}}\, ,
\end{equation*}
where $N_G$ is the number of reciprocal lattice vectors
$\vec{G}$ within an energy cut-off, and $\Omega_o$ is the volume
of the primitive cell. These bulk expansion coefficients
$B^{\sigma}_{n\vec{k}}(\vec{G})$ can be calculated by
diagonalizing the bulk Hamiltonian for each $\vec{k}$ point. In
our computational implementation, the bulk coefficients are
computed by employing the empirical pseudopotential method (EPM).\cite{cohen1988}

The total single-particle Hamiltonian of a nanostructure
with the kinetic energy and the ionic potential parts, where the
latter describes the atomistic environment within the
pseudopotential framework, is expressed by
\begin{equation*} \label{Hamiltonian}
\hat{H}=\hat{T}+\hat{V}_{PP}=-\frac{\hbar^2\nabla^2}{2m_o}+
\sum_{\sigma,\vec{R}_l,\alpha} W^{\sigma}_{\alpha}(\vec{R}_l)\,
\upsilon^{\sigma}_{\alpha} ( \vec{r}-\vec{R}_l-\vec{d}^{\sigma}_{\alpha}) \, ,
\end{equation*}
where $m_o$ is the free electron mass, the direct lattice vector
$\vec{R}_l$  gives the position of the primitive cell and
$\vec{d}^{\sigma}_{\alpha}$ specifies the relative coordinate of a
particular atom within the primitive cell. The weight function
$W^{\sigma}_{\alpha}(\vec{R}_l)$ keeps the information about the
composition of the system by taking values 0 or 1 depending on the
type of the atom located at the position
$\vec{R}_l+\vec{d}^{\sigma}_{\alpha}$.
$\upsilon^{\sigma}_{\alpha}$ is the local screened spherical
atomic pseudopotential of atom $\alpha$ of the material $\sigma$.

Hamiltonian matrix elements are evaluated within the basis set
\{$\vert n\vec{k}\sigma \rangle$\} that is
\{$\langle\vec{r} \vert n\vec{k}\sigma \rangle= \phi^{\sigma}_{n,\vec{k}}(\vec{r})$\} in the position representation.
The resulting Hamiltonian is diagonalized to yield {$C^{\sigma}_{n,\vec{k},j}$} coefficients.
The corresponding generalized eigenvalue problem is
\begin{equation*}\label{sch} 
\sum_{n,\vec{k},\sigma} \hspace{-3pt}  \langle n^{\prime}\vec{k}^{\prime}
\sigma^{\prime}\vert\hat{T}+\hat{V}_{\small PP}\vert n\vec{k} \sigma
\rangle \, C^{\sigma}_{n,\vec{k},j} \hspace{-2pt} = \hspace{-3pt} E_j \hspace{-3pt} \sum_{n,\vec{k},\sigma} \hspace{-3pt}
C^{\sigma}_{n,\vec{k},j} \langle
n^{\prime}\vec{k}^{\prime}\sigma^{\prime}\vert n\vec{k}\sigma
\rangle  \, ,
\end{equation*}
where the matrix elements are
\begin{equation*}\label{bulk11}
\langle n^{\prime}\vec{k}^{\prime}\sigma^{\prime}\vert
n\vec{k}\sigma \rangle = \delta_{\vec{k},\vec{k}^{\prime}}
\sum_{\vec{G}} B^{\sigma^{\prime}}_{n^{\prime}\vec{k}}(\vec{G})^*
B^{\sigma}_{n\vec{k}}(\vec{G}) \, ,
\end{equation*}
\begin{equation*}\label{Tmat1}
\langle n^{\prime}\vec{k}^{\prime}\sigma^{\prime} \vert \hat{T}
\vert n\vec{k}\sigma \rangle = \delta_{\vec{k},\vec{k}^{\prime}}
\sum_{\vec{G}} \frac{\hbar^2 \vert \vec{k}+\vec{G} \vert^2}{2m_o}
B^{\sigma^{\prime}}_{n^{\prime}\vec{k}}(\vec{G})^*
B^{\sigma}_{n\vec{k}}(\vec{G}) \, ,
\end{equation*}
\begin{align*}
\langle n^{\prime}\vec{k}^{\prime}\sigma^{\prime} \vert
\hat{V}_{PP} \vert n\vec{k}\sigma \rangle =
& \sum_{\vec{G},\vec{G}^{\prime}} 
B^{\sigma^{\prime}}_{n^{\prime}\vec{k}^{\prime}}
(\vec{G}^{\prime})^* B^{\sigma}_{n\vec{k}}(\vec{G}) \\ & \times
\sum_{{\sigma^{\prime\prime}},\alpha}
\mathcal{V}_\alpha^{{\sigma^{\prime\prime}}} (\vert
\vec{k}+\vec{G}-\vec{k}^{\prime} -\vec{G}^{\prime} \vert^2)
 \\ & \times  \mathcal{W}_\alpha^{{\sigma^{\prime\prime}}}(\vec{k}-\vec{k}^{\prime})
e^{-i(\vec{k}+\vec{G}-\vec{k}^{\prime} -\vec{G}^{\prime})
\cdot\vec{d}_\alpha^{{\sigma^{\prime\prime}}}} \, .
\end{align*}
Here $\mathcal{V}_\alpha^{{\sigma^{\prime\prime}}}$ and $\mathcal{W}_\alpha^{{\sigma^{\prime\prime}}}$
are the Fourier transformations of atomic pseudopotentials and weight functions, respectively.

In the case of dispersion-based approach,\cite{thesis} the separation of the $k$-space into $k_{\perp}$ and $k_{\parallel}$
(illustrated in Fig.~\ref{kgrid_fig}) leads to a separated eigenvalue problem of the form of
\begin{align*}
\sum_{n,\vec{k}_{\bot},\sigma} & \langle
n^{\prime}\vec{k}_{\bot}^{\prime}\vec{k}_{\parallel}\sigma^{\prime}
\vert \hat{T}+\hat{V}_{PP}\vert
n\vec{k}_{\bot}\vec{k}_{\parallel}\sigma \rangle \,
C^{\sigma}_{n,\vec{k}_{\bot},\vec{k}_{\parallel},j} \\ =
& E_j(\vec{k}_{\parallel}) \sum_{n,\vec{k}_{\bot},\sigma}
C^{\sigma}_{n,\vec{k}_{\bot},\vec{k}_{\parallel},j} \, \langle
n^{\prime}\vec{k}_{\bot}^{\prime}\vec{k}_{\parallel}
\sigma^{\prime} | n\vec{k}_{\bot}\vec{k}_{\parallel} \sigma \rangle \,,
\end{align*}
which is solved at definite $\vec{k}_{\parallel}$ values to obtain the
energy dispersion $E_j(\vec{k}_{\parallel})$.
This form does not put any restriction on the treatment of matrix
elements. One should only need to form a planar $k$-grid at a specific
$\vec{k}_{\parallel}$ point.\cite{thesis}

\vspace{-0.5cm}
\section*{Appendix B:  Oxide Passivation and Atomic Pseudopotentials}
\vspace{-0.3cm}

An important aspect of the electronic structure calculations is
the surface passivation. The surface of an unpassivated
nanostructure consists of dangling bonds which introduces surface
states lying in the otherwise forbidden band gap.
Surface passivation removes these surface states.
In our implementation, the surface passivation is provided by
embedding the NW structures into a host matrix of SiO$_2$.
However, the EPM calculations involving oxygen is rather nontrivial and, moreover,
the different lattice structure of
SiO$_2$, with reference to Si, causes strain effects.\cite{bulutay2007} To overcome
these obstacles, we introduce an \textit{artificial} monoatomic wide band gap
material that computationally substitutes silica. After
constructing the Si NW-based core, all the remaining crystal points
within the computational supercell are filled with the
\textit{artificial oxide atoms}. This artificial embedding
material has the same dielectric constant of SiO$_2$ and band edge
lineup, with respect to Si, however with the diamond structure to
prevent complications associated with strain\,\cite{bulutay2007}.

For the local empirical pseudopotentials of Si and SiO$_2$, we use
the functional form introduced by Freidel $et$
$al.$\,\cite{friedel1989} These authors suggest an analytical
expression to produce the pseudopotential form factor at a given
general wave number~$q$
\begin{equation}\label{form_facs}
V_{PP}(q)=\frac{a_1(q^2-a_2)}{e^{a_3(q^2-a_4)}+1} \left[
\frac{1}{2} \tanh\left( \frac{a_5-q^2}{a_6} \right) + \frac{1}{2} \right].
\end{equation}
We employ their parameters for Si (listed in
Table~\ref{EPM_ff}).\cite{friedel1989} On the other hand, we
have generated pseudopotential parameters\cite{thesis} for the
artificial SiO$_2$ (given in in
Table~\ref{EPM_ff}) which reproduces the experimental band
alignments of bulk Si/SiO$_2$ interface as 4.4~eV and 3.4~eV for
the valence~\cite{keister1999} and conduction~\cite{himpsel1988}
band edges, respectively. The units are arranged such that the
pseudopotential form factors come out in Rydbergs and the wave
number $q$ in Eq.\,(\ref{form_facs}) should be in the atomic units
(1/Bohr radius). For the plane-wave cut-off energies of EPM, we
use 14\,Ry for both Si and artificial SiO$_{\rm 2}$.

\begin{table}[t]
\caption{ \footnotesize Parameters of the pseudopotential form factors (see
Eq.\,(\ref{form_facs})) of Si and the wide band-gap matrix which
represents SiO$_2$. Units are given in the text.\label{EPM_ff}}
\begin{center}
\begin{tabular}{l c c c c c c}
\hline \hline
            & $a_1$ & $a_2$ & $a_3$ & $a_4$ & $a_5$ & $a_6$  \\
\hline
Si          & 106.0686 & 2.2278 & 0.606 & -1.972 & 5.0 & 0.3   \\
Matrix-Si   & 69.625 & 2.614   & 0.0786 & -19.1433  & 5.99 & 0.335  \\
\hline \hline
\end{tabular}
\end{center} \vspace{-0.5cm}

\end{table}

Though the surface chemistry of oxide-passivation requires more
elaborate first principles calculations, considering today's
computation limits, we believe that our approach offers a viable
alternative for the calculations of oxide-embedded large scale Si
nanostructures. Despite the missing surface relaxation and strain effects, the
reliability and competence of our computation method has gained
confidence in the context of embedded Si and Ge nanocrystals,\cite{bulutay2007}  and in
the energetics of single Si NWs and NW networks.\cite{keles}

The LCBB basis set is constructed by only employing the lowest two bulk conduction bands of
the core Si. We tested and observed that the further increase in the number of core bulk bands or
the inclusion of the bulk bands
of oxide-matrix do not alter the energetics and localization behavior of the states
within the energy window very close to band edge ($\sim$0.5 to 1~eV above the conduction band minimum).
Further discussion of LCBB method can be found in the original papers\cite{lcbb1997,lcbb1999} and
in our previous works.\cite{bulutay2007,keles}

\vspace{-0.3cm}


\begin{thebibliography}{55}
\bibitem [{not()}]{note1}%
  \BibitemOpen
  \href@noop {} {}\bibinfo {note} {We should note that there would be other
  interface states as well. However, as we only include the bulk silicon Bloch
  functions in the LCBB basis set, we do not probe most of such states.}\BibitemShut
  {Stop}%
\end{thebibliography}
\end{document}